\documentclass[twocolumn,showpacs,preprintnumbers,amsmath,amssymb]{revtex4}
\usepackage{tabularx,graphicx}\begin{document}
\newcommand{\beq}{\begin{equation}}
\newcommand{\eeq}{\end{equation}}
\newcommand{\beqn}{\begin{eqnarray}}
\newcommand{\eeqn}{\end{eqnarray}}
\newcommand{\bmath}{\begin{subequations}}
\newcommand{\emath}{\end{subequations}}

\title{ Superconductors as giant atoms predicted by the theory of hole superconductivity}
\author{J. E. Hirsch }
\address{Department of Physics, University of California, San Diego\\
La Jolla, CA 92093-0319}

\date{\today} 

\begin{abstract} 
The theory of hole superconductivity proposes that superconductivity originates in the fundamental electron-hole asymmetry of condensed matter and that it is an 'undressing' transition. Here we propose that a natural consequence of this theory is that
superconductors behave as giant atoms. The model predicts that the charge distribution in superconductors
is inhomogeneous, with higher concentration of negative charge near the surface. Some of this negative
charge will spill out, giving rise to a negative electron layer right outside the surface of the superconductor,
which should be experimentally detectable. Also superconductors should have a tendency to easily lose negative
charge and become positively charged.  Macroscopic spin currents are predicted to exist in superconducting bodies, giving rise to electric fields near
the surface of multiply connected superconductors that should be experimentally detectable.
\end{abstract}
\pacs{}
\maketitle 
 
\section{Motivation}

The presence of the discrete ionic potential in a metal causes the nature of the charge carriers in electronic
energy bands to change from electron-like to hole-like as the Fermi level rises from the bottom to the top
of the band. The theory of hole superconductivity\cite{hole1} proposes that superconductivity originates in a
fundamental asymmetry between electron and hole  carriers in metals: holes are always more dressed than electrons in a band\cite{hole2},
and this asymmetry is especially large for materials that are high temperature superconductors, for
materials reasons discussed elsewhere\cite{hole3}.  Superconductivity  is an 'undressing' transition\cite{hole4}, 
whereby dressed
hole carriers in the normal state undress upon pairing and become more free and more electron-like
in the superconducting state.

In (ideal) superconductors, electrons behave as if they were totally free. The conductivity is infinite and they
exhibit a perfect Meissner effect. As pointed out by London, such properties would be expected from a
big atom ('ein grosses diamagnetisches Atom')\cite{london}. Furthermore, experiments that measure the
London moment\cite{londonm} and the gyromagnetic effect\cite{gyro} show that the superfluid
carriers have the electronic charge (negative sign) and the $free$ electron mass\cite{elondon}.

In this paper we carry the principles suggested in the two paragraphs above to their logical ultimate consequence.
The reason that holes exist in the first place is due to the interaction of the electron with the discrete ionic 
lattice potential, and the holes are dressed because of electron-electron interactions in the presence of the
ionic potential. In other words, the effective mass of carriers in the normal state of metals is different from the
free electron mass both because of  electron-ion and  electron-electron interactions. In order for carriers
to become completely free in the superconducting state, as indicated by the above mentioned experiments,
they need to 'undress' from both of those interactions. We assume that is what happens to
the antibonding electrons at the top of the Fermi distribution   when a
metal goes superconducting.  This leads to a description of superconductors as giant atoms, and to some unusual and
interesting consequences. A preliminary discussion of the implications of charge imbalance in the
theory of hole superconductivity leading to a 'giant atom' picture was given in Ref. \cite{hole5}.

\section{The model}
The BCS theory of hole superconductivity\cite{holebcs} is not very different from conventional
BCS theory\cite{tinkham}. In BCS theory the Fermi distribution in a range of roughly $2\Delta$ around the Fermi energy gets modified by the development of pairing correlations. 
The same is true in the BCS theory of hole superconductivity, except that in addition by pairing
those carriers lower their effective mass and their kinetic energy as well as increase their
quasiparticle weight, i.e. they 'undress'. We assume that as a consequence of this
a fraction of roughly $2\Delta/\epsilon_F$ of the carriers become totally undressed from interactions with the discrete ionic charge distribution. Here $\epsilon_F$ is the Fermi
energy measured from the bottom of the electronic energy band, and
\beq
N_u\sim \frac{2\Delta}{\epsilon_F}N_e
\eeq
is the number of undressed electrons, with $N_e$ the total number of electrons in the band.

We consider a superconductor of spherical shape and assume that the undressed electrons
see a continuous positive
charge distribution of macroscopic dimensions, a macroscopic 'Thomson atom '\cite{thomson} of radius $R$, of total charge $Ze$ uniformly distributed over the sphere. The value of $Z$ will be discussed later. The electric field at position $\vec{r}$ from the center seen by the
undressed electrons is
\beqn
\vec{E}(\vec{r})&=&\frac{Ze\vec{r}}{R^3}  \ \ \ \ \ \  r<R  \nonumber \\
&=& \frac{Ze}{r^2} \ \ \ \ \ \  r>R
\eeqn
and the electronic potential energy is, with $U_0=-3Ze^2/2R$
\beqn
U(r)&=&U_0+\frac{1}{2}\frac{Ze^2}{R^3} r^2  \ \ \ \ \ \   r<R  \nonumber \\
&=& -\frac{Ze^2}{r}    \ \ \ \ \ \ \ \ \ \ \ \ \ \ \ \   r>R
\eeqn
defined so that $U(r=\infty)=0$. Thus we have a harmonic oscillator potential inside the sphere and
the ordinary Coulomb potential outside.

\section{Bohr orbits}
We follow the treatment of the Bohr atom for the model Eq. (2). Assume the electron moves in a circular orbit of radius $r$, 
with energy
\beq
E(r)=\frac{1}{2}mv^2+U(r)
\eeq
For the motion to be stable the centripetal acceleration equals the electrostatic attraction
\beq
\frac{mv^2}{r}=\frac{Ze^2r}{R^3}
\eeq
so that
\bmath
\beq
v(r)=\omega r
\eeq
\beq
\omega=(\frac{Ze^2}{mR^3})^{1/2}
\eeq
\emath
Note that this is a 'rigid rotation' with angular velocity $\omega$.

As in the Bohr atom, the orbital angular momentum will be quantized
\beq
L=mvr=n\hbar
\eeq
which leads to a quantization of the orbits
\bmath
\beq
r_n=r_0\sqrt{n}
\eeq
\beq
r_0=(\frac{R^3 a_0}{Z})^{1/4}
\eeq
\emath
with $a_0=\hbar ^2/me^2$ the ordinary Bohr radius. The radius of the orbit increases much more slowly than in 
the Bohr atom. The energy of an electron in the $n$-th orbit is
\beq
\epsilon_n=\hbar\omega n +U_0
\eeq
When $r_n$ exceeds the radius $R$, Eq. (8) are replaced by the ones of the ordinary Bohr atom,
$r_n=r_{0B}n^2$, $r_{0B}=a_0/Z$. Figure 1 shows a picture for a body that can accommodate
6 orbits in its interior.
\begin{figure}
\resizebox{6.5cm}{!}{\includegraphics[width=7cm]{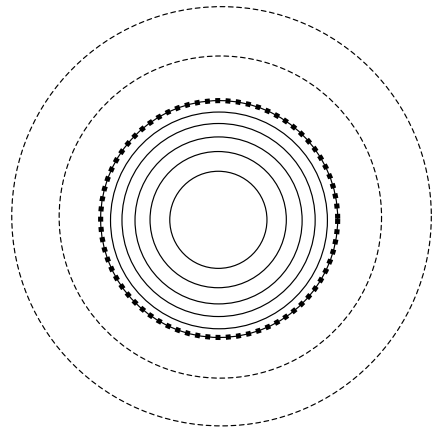}}
\caption{Single electron Bohr orbits for a giant atom that can accommodate 6 orbits in its interior. The surface is denoted 
by the dotted line. The orbits outside (dashed lines) are ordinary Bohr orbits. }
\label{atom1}
\end{figure}

To determine the number of electrons that can fit in each orbit we resort to the
solution of the three-dimensional quantum harmonic oscillator\cite{messiah}. The
eigenenergies are given by Eq. (9), and the
degeneracy of the $n$-th energy level (per spin) is
\beq
C_n=\frac{(n+1)(n+2)}{2}
\eeq
so that approximately (for large $n$) the number of electrons of both spins in the
$n$-th orbit is simply
\beq
g_n=n^2
\eeq
which is one-half of that  in the ordinary Bohr atom.

Because of the large degeneracy Eq. (11) and the fact that the radius of the orbit increases
slowly with $n$  (Eq. (8)), the undressed electrons will pile up in the orbits with large $n$. We can
estimate the dependence of the undressed electron density with $r$. The spacing between two
neighboring orbits is
\beq
r_{n+1}-r_n\sim \frac{r_0}{2\sqrt{n}}=\frac{r_0^2}{2r_n}
\eeq
hence the density of undressed electrons at $r$ is
\beq
n_u(r)\sim \frac{n^2}{4\pi r_n^2 (r_{n+1}-r_n)}=\frac{r^3}{2\pi r_0^6}
\eeq
and it increases rapidly with $r$. We conclude that the superfluid undressed electrons are pushed out of
the bulk of the superconductor towards the surface, as discussed in Ref. \cite{hole5}.
The enhanced charge density near the surface can also be understood from the fact that a harmonic oscillator spends most of the time near the region of maximum elongation.

In this simple one-electron picture, the quantum number of the orbit at the surface $n_{max}$, corresponding to
$r_{nmax}=R$ is given by
\beq
n_{max}=(\frac{ZR}{a_0})^{1/2} .
\eeq
The number of electrons that   fit in all orbits up to $n_{max}$ is
\beq
N_f=\sum_0^{n_{max}} 2C_n \sim \frac{n_{max}^3}{3}=\frac{1}{3} (\frac{ZR}{a_0})^{3/2}
\eeq
It is interesting to note that (for fixed $Z$) $N_f$ increases as $R^{3/2}$. On the other hand,
we expect that the number $N_u$ of electrons that undress Eq. (1) is proportional to the volume, i.e. $R^3$.
This suggests that some of the undressed electrons will 'spill over' the surface and occupy orbits
$outside$ the superconducting body. To obtain an estimate of this effect we need to take
into account the effect of screening.

\section{Screening and spill-over}
According to Eq. (15), the number of electrons that fit inside the sphere is much larger than $Z$. If $Z=N_u$, the number
of undressed electrons, this would indicate that the undressed electrons do not reach the surface. This is
however incorrect, because as the inner orbits become filled the effective charge seen by the electrons
in the outer orbits becomes much smaller than the bare $Z$ due to screening by the electrons in the
inner orbits.

As a simple approximation we modify Eq. (8) to read
\bmath
\beq
r_n=r_{0n}\sqrt{n}
\eeq
\beq
r_{0n}=(\frac{R^3a_0}{Z_n})^{1/4}
\eeq
\emath
and take as the effective charge $Z_n$ the bare charge $Z$ minus the charge of the electrons in orbits
smaller than $n$, i.e.
\beq
Z_n\sim Z-\frac{n^3}{3}
\eeq
Calling again $n_{max}$ the quantum number for the orbit corresponding to $r_{nmax}=R$, it satisfies
\beq
n_{max}=[\frac{R}{a_0}(Z-\frac{n_{max}^3}{3})]^{1/2}
\eeq
and 
\beq
N_{spill}=Z_{nmax}
\eeq
is the number of electrons that 'spill over' the surface. If $N_{spill}<<Z$ (as confirmed below)
we have approximately
$n_{max}=(3Z)^{1/3}$ and Eq. (18) yields
\beq
N_{spill}\sim \frac{2a_0}{R}Z^{2/3}
\eeq
as the number of electrons that are expelled  from the superconductor. Since $Z=N_u\propto R^3$, $N_{spill}$ increases
linearly with $R$. For a superconductor of $1cm$ radius, $N_{spill}$ could be of order $10^8$ electrons.

Once outside the superconductor the electrons will occupy the lowest Bohr orbit available to them near
the surface, which certainly has enough degeneracy to accommodate all $N_{spill}$. Because they are
outside the superconductor these electrons should be easily removed by contact or friction.
Consequently it should be much easier for a metal to lose electrons by contact with other bodies if it is in
the superconducting state,   and superconductors should quite generally be at
the top of the triboelectric series\cite{tribo}. A qualitative picture of superconductors as it emerges from
the physics discussed here is shown in Fig. 2.
\begin{figure}
\resizebox{6.5cm}{!}{\includegraphics[width=7cm]{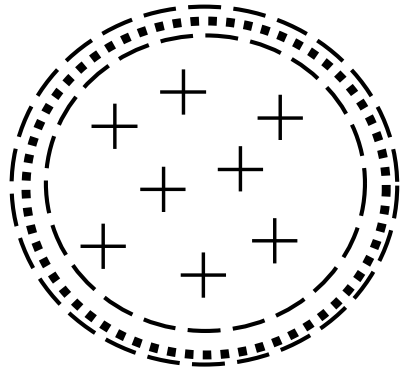}}
\caption{Schematic picture of a spherical superconductor. Negative charge is expelled from the bulk
to the surface. The  surface is denoted by the dotted line. A layer of negative charge exists outside the surface. }
\label{atom2}
\end{figure}

Furthermore, the electronic layer outside the surface is likely to affect the friction properties of the superconductor,
by providing a 'lubricating layer' on top of which another material would slide. As a matter of fact,
an abrupt drop in sliding friction between a lead surface and solid nitrogen has been observed when
$Pb$ enters the superconducting state\cite{friction}. We infer that this effect is due to the physics of the 
electron layer outside the surface discussed here, and hence that a friction drop should be observed for all superconductors.

We predict that this electronic layer outside the surface should exist for all $simply$ $connected$ superconductors. It may be possible to confirm this  
by direct observation through sensitive optical detection techniques. The surface will become
more 'fuzzy' when the metal enters the superconducting state.

\section{Spin currents}

\begin{figure}
\resizebox{6.5cm}{!}{\includegraphics[width=7cm]{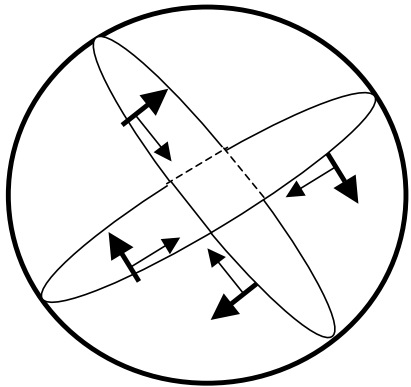}}
\caption{Macroscopic spin currents in a spherical superconductor.  The arrow perpendicular to the orbit indicates
the direction of the electron magnetic moment (opposite to the electron spin), and the arrow parallel to the orbit indicates
the direction of motion. }
\label{atom3}
\end{figure}

In a giant atom there will be a large spin-orbit coupling. It is then not surprising that the picture discussed here
will give rise to macroscopic spin currents flowing in the superconductor. The qualitative picture is shown in
Figure 3. The orbits depicted have lower energy than the corresponding ones obtained reversing
either the velocity direction or the magnetic moment direction. In this state, parity is broken but 
time-reversal invariance is preserved\cite{spinsplit}.

First we give a heuristic argument for why spin currents will develop when a material goes superconducting.
As discussed in the previous sections, negative charge, i.e. electrons, are expelled from the interior
towards the surface. Electrons carry spin and an associated magnetic moment $\vec{m}$. As discussed in
Ref. \cite{hallferro}, a magnetic moment moving with velocity $\vec{v}$ generates an electric
dipole in the laboratory frame
\beq
\vec{p}=\gamma \frac{\vec{v}}{c} \times \vec{m}
\eeq
A force will be exerted on this dipole due to the positive background, which will deflect it in direction $-\vec{p}$, as shown
schematically in Fig. 4. This physics was discussed in Ref. \cite{hallferro} in connection with the anomalous
Hall effect in ferromagnetic metals. When electrons move towards the surface, up and down spin electrons are
deflected in opposite directions perpendicular to $\vec{m}$ and to the radial velocity, building up the spin current.
\begin{figure}
\resizebox{6.5cm}{!}{\includegraphics[width=7cm]{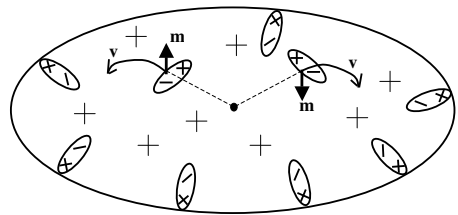}}
\caption{Electrons are expelled towards the surface when the system goes superconducting, and are
deflected by the interaction of the moving magnetic moment (equivalent to an electric dipole moment) with
the positive background, giving rise to a spin current. The figure also shows the resulting equivalent dipole
configuration near the surface.
}
\label{atom4}
\end{figure}

We can discuss this effect using the same analysis as for atoms. In the rest frame of the orbiting electron, the
electric field gives rise to a magnetic field that couples to the magnetic moment of the electron and gives
rise to the spin-orbit interaction energy. Including the correction for Thomas precession,\cite{baym}
\beq
U_{so}=\frac{e}{2mc^2}\vec{S}\dot(\vec{v}\times\vec{E})
\eeq
with $\vec{S}$ the electron spin. This yields for our model
\beq
U_{so}=\frac{\omega^2}{2mc^2} \vec{L}\cdot  \vec{S}
\eeq
so that the energy is minimized when $\vec{L}$ is antiparallel to $\vec{S}$, or
$\vec{L}$ is parallel to $\vec{M}$, as shown in Fig. 3. The magnitude of the spin-orbit coupling
increases as the orbit approaches the surface.

A spin current flowing in a solid will give rise to an electric polarization and an associated electric
field originating in the elementary electric dipoles Eq. (21). In this way electric fields can arise
in and around a solid in the absence of electric charges, charge currents and magnetic fields. This unusual physical
effect has been discussed in the past in connection with a proposed symmetry-broken state of the metal Cr that
would sustain a spin current\cite{chrom,chrom2}, and in connection with a possible  state of
aromatic molecules that  may exhibit a spin current\cite{aromatic}. Recently the equations governing
electric fields from spin currents and the resulting  electric
field patterns have been analyzed   in connection with possible 
spintronics applications \cite{spincurrent}.

The electric field generated by a spin current in a simple slab geometry is\cite{chrom2}
\bmath
\beq
\vec{E}=\frac{4\pi}{c}\mu_B j_{spin}
\eeq
\beq
j_{spin}=n_\uparrow v_\uparrow-n_\downarrow v_\downarrow
\eeq
\emath
with $n_\sigma$ the density of electrons of spin $\sigma$ and $v_\sigma$ their velocity,
and $\mu_B$ the Bohr magneton. 
From Eq. (6) with $R=1cm$ and $Z=1$\cite{explain} we obtain a velocity at the surface $v\sim16,000cm/s$.
For $\Delta / \epsilon_F \sim 10^{-3}$ and assuming $n_\sigma  \sim 10^{23}/cm^3$
, we obtain as an estimate for the electric field near the
surface
\beq
E \sim 20 mV/mm  
\eeq
This is a  large field,  easily measurable. We emphasize however that this estimate is very rough.
Figure 5 shows an example of a superconducting ring with spin currents and the associated
electric field pattern. An electric field would  not appear for a simply connected superconductor. Far
from the superconductor the electric field decays rapidly as the fourth power of the distance 
since it is quadrupolar in nature.
\begin{figure}
\resizebox{6.5cm}{!}{\includegraphics[width=7cm]{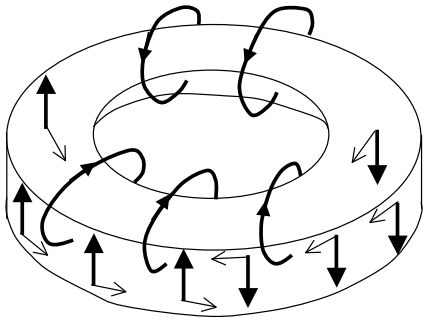}}
\caption{Pattern of electric field (curved lines) expected to arise from a superconducting disk with
a cavity. The vertical arrows denote the direction of the magnetic moment of the superelectrons and
the horizontal arrows the direction of their motion. No magnetic field nor charge current is present.}
\label{atom5}
\end{figure}

Detection of such electric fields originating in spin currents in superconductors by direct measurement
should be experimentally possible, although care must be taken to avoid extraneous contributions from 
stray charges. Another manifestation of these electric fields is that a force should exist between two superconducting rings, in the absence of magnetic fields and currents, as shown in Figure 6: depending on
their relative orientation the force will be attractive or repulsive. This force should be
small but measurable. 
We expect these spin currents to exist in the interior of all  superconductors as well as near
the surface, in the absence of charge currents and magnetic fields. The spin currents near the surface 
will be modified but persist when
supercurrent flows.

\begin{figure}
\resizebox{6.5cm}{!}{\includegraphics[width=7cm]{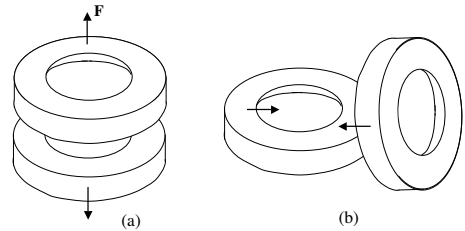}}
\caption{Force between two superconducting rings due to the electric fields produced by the spin currents.
Depending on their relative orientation the force can be repulsive (a) or attractive (b). The force should
exist in the absence of charge currents and magnetic fields.}
\label{atom6}
\end{figure}

There should be other experimental ways to detect the existence of spin currents in 
superconductors. For example, spin-polarized neutron scattering in an adequate geometry
should be able to detect differences in cross sections for different neutron polarization which
would not be expected in the absence of spin currents. Another possibility would be 
photoelectron spectroscopy with circularly polarized photons. This approach has been recently
used\cite{photovarma} in connection with a proposed state of high temperature superconductors that breaks
time reversal invariance\cite{varma}. We point out that even though the state discussed here (Fig. 3)
does not break time reversal invariance, that experiment should also be sensitive to
the symmetry breaking of interest here. These and other experiments should be
analyzed in detail.

\section{Some qualitative considerations}
As discussed by London\cite{london}, a 'giant atom' of dimensions larger than the London
penetration depth will exhibit a Meissner effect. Thus the model discussed here in a sense
justifies the London equation. It is also interesting to note that the present picture allows
for an understanding of the Meissner effect from the point of view of a 'perfect conductor'.
Recall that in the usual understanding a perfect conductor would not expel magnetic flux
if the magnetic field preexists before the perfectly conducting state is attained. However,
if charge is expelled from the interior when the system goes superconducting as
described here, the Lorenz force on the radially moving charge will deflect it so as to 
set up circulating currents that will screen the magnetic field in the superconductor.

Furthermore the present picture allows for an intuitive understanding of the 'London moment'\cite{elondon},
the observation that magnetic fields exist in the interior of rotating superconductors.
The conventional description is that when a superconductor is set into rotation the superfluid
electrons 'lag behind' near the surface due to inertia and zero viscosity, thus creating a current and
a magnetic field in the interior. However such description does not make sense when the system 
is rotating in the normal state and is subsequently cooled to become superconducting. In the
description here the lagging supercurrent at the surface will naturally arise because the
superfluid that is expelled from the interior will experience a Coriolis force when moving radially
outwards.

Also one may ask the question whether a magnetic field would be observed if the superconductor is
at rest and the observer is rotating. In the usual understanding no magnetic field is expected. In the present
description however a magnetic field would be observed due to the existing spin currents: the spins
moving in direction opposite to the observer will be Lorenz-contracted relative to the opposite spin,
giving rise to a magnetic field. This then extends the principle of relativity to rotating superconductors.

Finally we note that the existence of spin currents in superconductors predicted here is naturally
expected in a BCS description. Namely, a Cooper pair $c_{k\uparrow}^\dagger c_{-k\downarrow}^\dagger$
carries a spin current (i.e. spin up moves in the $k$, spin down in the $-k$ direction). It is only
because in the ordinary BCS treatment $k$ and $-k$ are equally occupied by Cooper pairs that no
spin currents were found before. The description of the physics discussed in this paper in a 
BCS framework will be the subject of future work.

\section{Discussion}

Many different theories of superconductivity exist and it is difficult to ascertain which describes reality,
since they often propose competing but seemingly plausible explanations for known experimental facts. For
that reason it is useful to spell out predictions of theoretical frameworks $before$ experiments to
test these predictions are performed. We summarize here the predictions of the theory of hole
superconductivity in connection with the physics discussed in this paper. Except for the ones
noted, none of these predictions has yet been experimentally tested, and to our knowledge no
other theory has made similar predictions thus far.

1) The superfluid carriers in all superconductors are bare electrons. Unfortunately this particular one is
a 'postdiction', verified in many superconductors long ago\cite{londonm,gyro}.

2) Superconductors have a tendency to expel negative charge from their interior. As a consequence,
the charge distribution inside simply connected
superconductors is inhomogeneous, with more positive charge deep
in the interior and more negative charge close to the surface.

3) A layer of negative charge should exist just outside the surface of any simply connected superconductor. 
Hopefully this should be verifiable by direct observation. This should
lead to an anomalous drop in sliding friction between the surface of a superconductor and a non-superconductor,
and an even larger drop in friction between the surfaces of two superconductors, as the
superconducting state sets in. This friction drop has  so far been observed in a single superconductor (Pb) 
  with another non-superconducting body\cite{friction}. It is not known experimentally whether the effect will exist
for other superconductors nor for friction between two superconductors.

4) A superconductor should easily lose electrons by contact or friction with a non-superconducting body.

5) Consequences of charge expulsion physics in the mixed state and for the properties of grain boundaries
were discussed in ref. \cite{hole5}.

6) Macroscopic spin currents should exist in all superconductors. They should lead to observable 
consequences, in particular electric  fields should arise near the surface of superconducting
rings (or any topology that can sustain persistent currents) in the
absence of electric charges, magnetic fields and electric currents.

7) The existence of these electric fields  should give rise to attractive
or repulsive forces between two superconducting rings depending on their relative
orientation, in the absence of magnetic fields and charge currents. 

(8) Differences in
cross sections in scattering experiments that are sensitive to the parity broken ground state of
superconductors  such as spin polarized neutron scattering or
photoelectron spectroscopy with circularly polarized light should be observed, resulting from the presence of spin currents. 

(9) An observer rotating on top of a superconductor with angular velocity $\vec{\omega}$
will measure a magnetic field that corresponds to a magnetic field in the interior of the superconductor
\beq
\vec{B}=\frac{2mc}{e}\vec{\omega},
\eeq
the same that is measured if the observer is at rest and the superconductor is rotating with angular velocity
$-\vec{\omega}$.
 This magnetic field would not be observed in the absence of spin currents in the superconductor.

In connection with the existence of forces between superconductors (point 7) above)
the possibility of Van der Waals forces also comes to mind, which should be especially
important in the case of spherical superconductors. Such forces may play a role in
the remarkable experiments of Tao et al\cite{tao}.

The physics discussed in this paper reflects the fundamental asymmetry of positive and negative charge
that is the foundation of the theory of hole superconductivity. To the extent that all superconductors exhibit
this charge asymmetry the physics proposed by the theory of hole superconductivity is validated.
If a single superconductor were not to show this physics it would disprove the fundamental principle
on which the theory is based.

In the normal state of a metal there is no experiment that can determine the sign of the elementary charge
carriers without extracting the electron from the metal\cite{thomson2}. In the superconducting state instead there are
experiments that show that the superfluid carriers inside the superconductor have negative 
charge\cite{elondon}. Moreover, the results in this paper predict that superconductors make the fundamental
charge asymmetry of matter extraordinarily apparent by pushing the negative electrons $outside$ the surface
of the superconducting body.

There exist many classes of materials that become superconducting and exhibit a variety of different
properties in the normal state, and different theories have been proposed to explain each of them. It is common
to emphasize that the more important problem is to explain the normal state properties. Instead, the
theory of hole superconductivity focuses on what is common to all superconductors which it predicts to show
up crystal-clearly in the superconducting state as discussed in this paper.  If the theory of hole superconductivity
is correct, cuprates are not d-wave superconductors, heavy fermion materials are not d-wave nor p-wave
superconductors, $Sr_2RuO_4$ is not a p-wave superconductor, organic superconductors are not unconventional,
and conventional superconductors and $MgB_2$ are not electron-phonon-driven. Instead,
superconductivity is a manifestation of the fundamental charge asymmetry of matter, namely that positive protons
are 2000 times heavier than negative electrons. All superconductors exhibit the same essential
physics and are driven superconducting by the same physical principle,  hole undressing, which
converts dressed holes into undressed electrons. Confirmation of these predictions and usage of the criteria
derived from the theory of hole superconductivity that favor higher $T_c$'s should pave the way for 
finding new and better superconducting materials.


\begin{references}
\bibitem{explain} We assume that an electron in the outer orbit 'sees' $Z=1$ because the rest of the positive
charge has been screened by the other electrons.  This is clearly a very crude approximation.
\end{references}
 \end{document}